\documentclass{cta-author}

{}
{}
{}
\usepackage[frozencache]{minted}
\usepackage[export]{adjustbox}
\usepackage{wrapfig}
\usepackage{array}
\usepackage{booktabs}
\usepackage{url}
\usepackage[normalem]{ulem}
\usepackage[flushleft]{threeparttable}
\usepackage{tcolorbox}
\nolinenumbers
\usepackage{color}
\definecolor{dkgreen}{rgb}{0,0.6,0}
\definecolor{gray}{rgb}{0.5,0.5,0.5}
\definecolor{mauve}{rgb}{0.58,0,0.82}

\begin{document}

\supertitle{Submitted to IET Software Journal}

\title{ConsiDroid: A Concolic-based Tool for Detecting SQL Injection Vulnerability in Android Apps}

\author{\au{Ehsan Edalat$^{1}$}, \au{Babak Sadeghiyan$^{1\corr}$}, \au{Fatemeh Ghassemi$^{2}$}}

\address{\add{1}{Amirkabir University of Technology}
\add{2}{University of Tehran}
\email{basadegh@aut.ac.ir}}

\begin{abstract}
In this paper, we present a concolic execution technique for detecting SQL injection vulnerabilities in Android apps, with a new tool we called ConsiDroid. We extend the source code of apps with mocking technique, such that the execution of original source code is not affected. The extended source code can be treated as Java applications and may be executed by SPF with concolic execution. We automatically produce a DummyMain class out of static analysis such that the essential functions are called sequentially and, the events leading to vulnerable functions are triggered.

We extend SPF with taint analysis in ConsiDroid. For making taint analysis possible, we introduce a new technique of symbolic mock classes in order to ease the propagation of tainted values in the code. An SQL injection vulnerability is detected through receiving a tainted value by a vulnerable function. Besides, ConsiDroid takes advantage of static analysis to adjust SPF in order to inspect only suspicious paths. To illustrate the applicability of ConsiDroid, we have inspected randomly selected 140 apps from F-Droid repository. From these apps, we found three apps vulnerable to SQL injection. To verify their vulnerability, we analyzed the apps manually based on ConsiDroid’s reports by using Robolectric.
\end{abstract}

\maketitle

\section{Introduction}\label{intro}
At the end of 2017, 2.7 million devices used the Android operating system \cite{goooglePlayStat}. It makes 88 percent of total mobile devices \cite{osStat}. Also, there were 3.3 million apps available on Google Play in March 2018. These numbers show that Android is the most popular OS among users and developers. Hence, it is a favorite target for attackers in order to exploit security vulnerabilities on Android apps to affect a large number of victims. For preventing such attacks, various approaches are presented. There exists a set of approaches such as  \cite{chex,APSET,vulhunter,staticCommunication,smith2014injection} based on static analysis to detect security vulnerabilities. These approaches have inherited the natural weakness of static analysis, i.e., high false alarm rates and the inability of analyzing dynamic code loaded at runtime.

Dynamic analysis is another approach, which overcome the static analysis weakness. In this approach, unlike static analysis, the app is analyzed when it is running. To run apps, the input data and events that trigger their various parts should be generated artificially. There are two main techniques used in dynamic approaches for input generation, i.e., fuzzing and concolic input generation \cite{motherOfSurvays}. Fuzzing generates the inputs randomly. However, this technique suffers from low code coverage. For instance, Android Monkey \cite{monkey} is a fuzzer for Android apps. Concolic input generation technique supports high code coverage, and hence, more states of the program are inspected during the analysis (see Section \ref{concolic}). This technique improves the symbolic analysis technique, by which, variables are represented symbolically, and their values are chosen by a constraint solver to cover more execution paths of a program. In this technique, a program is executed with symbolic and concrete input values simultaneously to resolve cases when the constraint solver cannot generate new inputs for symbolic values. In this paper, we choose concolic executions to analysis more lines of code. It worth mentioning that there is not any concolic and even symbolic engine for testing Android apps. Therefore, we are going to use and extend SPF\footnote{Symbolic Path Finder} \cite{spf}, a symbolic extension of Java Pathfinder, in order to test and detect vulnerabilities in Android apps by using the concolic testing technique. In this paper, we present a tool which we call ConsiDroid, an abbreviation for \textbf{Con}colic-based \textbf{S}QL \textbf{I}njection vulnerability detection tool in An\textbf{Droid} Apps.

For concolic execution of an Android app by SPF, we extend apps under the test to make them behave like Java programs. Android apps have no ``main'' function as their entry point. They have different ways of starting to run, such as tapping a URL or receiving an incoming call. Without changing the source code, we extend the test app by adding new classes, which indicate how to start running the app. Such new classes were already introduced in \cite{sigdroid}, which we call them DummyMain classes in this paper. They are produced through static analysis of the app. Unlike Sig-Droid, we produce them in such a way that only events leading to call a vulnerable function are considered. Our aim of such a method is to overcome the path explosion problem of concolic execution. Google provides an SDK\footnote{Software Development Kit} to ease the development of Android apps. These apps are built by extending the Google SDK, which turns to an inseparable part of the apps. Concolic execution of Android apps in SPF is impossible without the existence of SDK classes, which we trust in. Presence of SDK classes during concolic executions lead to path divergence problem. Hence, we use the mock class idea \cite{sigdroid} to emulate SDK functionality. Mock classes are the same as real ones except that the bodies of their class functions have been omitted. Also, their return values have been replaced with default ones.

The quality of an app code is a category in OWASP Mobile Top 10 \cite{owaspMobile}. Our focus is SQL injection vulnerability as it is directly related to the quality of code. This attack happens when the inputs fed into SQL related functions are not adequately controlled. Nonetheless, there exist other related vulnerabilities which our approach is suitable for them, like OS shell injection. We also modify and extend the concolic engine in SPF to optimally detect SQL injection vulnerabilities. For detecting this kind of vulnerabilities, we use taint analysis, which tracks values from sources to sinks. We perform a dynamic taint analysis through concolic execution with the help of the symbolic mock classes idea. Besides, we take advantage of static analysis for enhancing the performance of ConsiDroid. To this aim, we automatically extract vulnerable paths through static analysis and give them precedence during concolic execution. With this idea, the time analysis reduces considerably. To evaluate our technique, besides manually developed apps, we randomly selected 140 real-world apps from F-Droid \cite{fdroid}, which is an open source repository for Android apps. As a result, ConsiDroid could detect three vulnerable apps correctly. Also, we observe that the increase in time and code coverage of our dynamic analysis are smaller than the increase in test apps complexity.

The remainder of the paper is organized as follows. Section \ref{background} covers some background about concolic execution, Android OS, challenges of concolic execution in Android apps, taint analysis, and SQL injection vulnerability. Section \ref{Overview} gives an overview of ConsiDroid. Section \ref{staticAnalysis} describes how to produce DummyMain classes and vulnerable paths. Section \ref{mock} gives information about mock and symbolic mock classes. Section \ref{extendedConcolic} presents the ideas that are used in extended concolic execution for detecting SQL injection vulnerability. To validate the results of ConsiDroid, we manually explore the derived vulnerable paths by our tool with the help of Robolectric tool in Section \ref{robo}. To illustrate the applicability of ConsiDroid in real-world apps, we apply it to several applications in Section \ref{evaluation}. Section \ref{relatedWork} is about related work and the paper is concluded in Section \ref{conclusion}.

\section{Background}\label{background}
In this part, we introduce concolic execution and its pros and cons briefly. Besides, we explain some essential topics in Android. We discuss challenges when dealing with concolic testing of Android apps. We explain taint analysis as we use it to track the flow of values to detect SQL injection. Finally, we present the conditions that an attacker needs to exploit for SQL injection.
\begin{figure*}
	\begin{tabular}{p{0.5\textwidth}p{0.5\textwidth}}
		\begin{minipage}{.50\textwidth}
			\centering
			\includegraphics[scale=0.75, left]{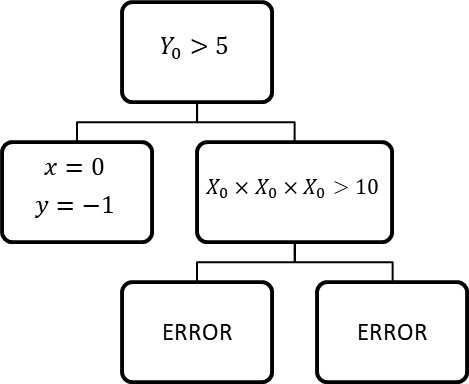}
			\caption{Execution tree for listing \ref{simplecode}; the left and right child nodes indicate the {\it else} and {\it then} branches, respectively.}
			\label{fig:exeTree}
		\end{minipage}
		&
		\noindent\begin{minipage}{.43\textwidth}
			\begin{listing}[H]
				\caption{Simple code in C for testing with concolic execution.}\label{simplecode}
				\begin{minted}[mathescape,
				fontsize=\scriptsize,
				linenos,
				numbersep=-5pt,
				gobble=2,
				frame=lines,
				tabsize=2,
				breaklines,
				obeytabs=true
				framesep=2mm]{c}
				void testMe(int x, int y){
					if(y>5){
						if(x*x*x > 10){
							assert(false);
						}
					}
				}
				void main (){
					int x=symbolic_input();
					int y=symbolic_input();
					testMe(x,y);
				}
				\end{minted}
			\end{listing}
		\end{minipage}\hfill
	\end{tabular}
\end{figure*}
\subsection{Concolic Execution}\label{concolic}
Concolic execution means simultaneously executing a program with symbolic and concrete values. In symbolic execution, we consider some values as symbolic. For instance, the simple code in Listing \ref{simplecode} is executed with the symbolic variables $X_{0}$ for $x$ (line 9) and $Y_{0}$ for  $y$ (line 10). In this approach, conditions are essential points. Path conditions are logical conjunction of conditions, which are collected during the execution of a specific path. For example, in testMe function, one path condition can be $(Y_{0}>5 \wedge X_{0}\times X_{0}\times X_{0}\leq10)$. A path condition indicates which branches of the code have been executed recently. The execution tree of a program is the mixture of all path conditions (Fig. \ref{fig:exeTree}). The intermediate nodes contain constraints while the leaf nodes contain the real values satisfying the constraints over the path. The constraint solver generates these constraints. In symbolic execution, we want to cover all possible paths in the execution tree. Therefore, we execute the code and construct a path condition during the execution. For each conditional statement, symbolic execution is forked to cover both \emph{then} and \emph{else} branches. Finally, the execution tree of the program is built. For producing a test input for a specific path in the execution tree, we use a constraint solver like SMT, which solves logical expressions and produces suitable values. Constraint solvers have some weakness in solving some complex expressions like non-linear expressions, e. g., $(X_{0}\times X_{0}\times X_{0}\leq10)$. In pure symbolic execution, testing of the code is stopped at this point, and we are not able to produce test inputs to execute some lines of code like line 4.

For alleviating the problem, concolic execution is used, which executes the code with concrete values in addition to symbolic ones. First, concrete values are generated randomly. Assume that $x=0$ and $y=-1$. These values produce the condition $(PC_{1}=Y_{0}\leq 5)$. New concrete values are generated by negating the last constraint in the path condition $(\neg (Y_{0}\leq5)=(Y_{0}>5))$. A constraint solver can solve the condition $(Y_{0}>5)$, resulting in $x=0$ and $y=6$ as the new values. These new values build the path condition $(PC_{2}=Y_{0}>5 \wedge X_{0}\times X_{0}\times X_{0}\le 10)$. Whenever a constraint solver cannot solve an expression, concrete values are randomly generated ($y=6$ and $x=11$). With this trick, we may reach points in code that cannot be explored during symbolic execution (line 4). So, the concolic execution has better code coverage than pure symbolic execution.

With concolic execution, the problems of static analysis, i.e., the existence of false alarms\cite{dart} and the ignorance of code loaded at runtime \cite{motherOfSurvays}, are handled as the code are executed. However, concolic execution has some issues:
\begin{itemize}
	\item \textbf{Path explosion}: When we analyze real-world programs, we face too many lines of code. Hence, their execution trees contain a large number of paths. Testing in this situation is time and memory consuming. We tackle this problem by presenting our hybrid concolic execution for testing Android apps. In this approach, we take advantage of static analysis to make concolic execution targeted.
	\item \textbf{Frameworks and environment modeling}: Many programs are developed with frameworks. It means that a developer uses third-party libraries and software kits. In our testing approach, we trust this kind of code, and we only test developers’ program. Also, in different systems, we face a different kind of environmental issues. For example, in the Android system, apps have event-driven nature, and we should model these events. In this work, we use the idea of mock classes \cite{sigdroid}. Also, we present symbolic mock classes as our idea in order to overcome these issues.
\end{itemize}
Furthermore, we use SPF\cite{spf} as a concolic engine, and we extend it in order to detect SQL injection vulnerability in Android apps. SPF is built on JPF\cite{jpf}, a Java bytecode model checking tool. In SPF, bytecode is converted to 3-address intermediate instructions. These instructions execute on a modified JVM. SPF supports different kind of constraint solvers, including solvers which support strings. Also, in SPF, we can indicate which variables or methods should be symbolic.

\subsection{Challenges on Concolic Execution in Android Apps}
\label{android}
\begin{figure}
	\begin{center}
		\includegraphics[scale=0.4]{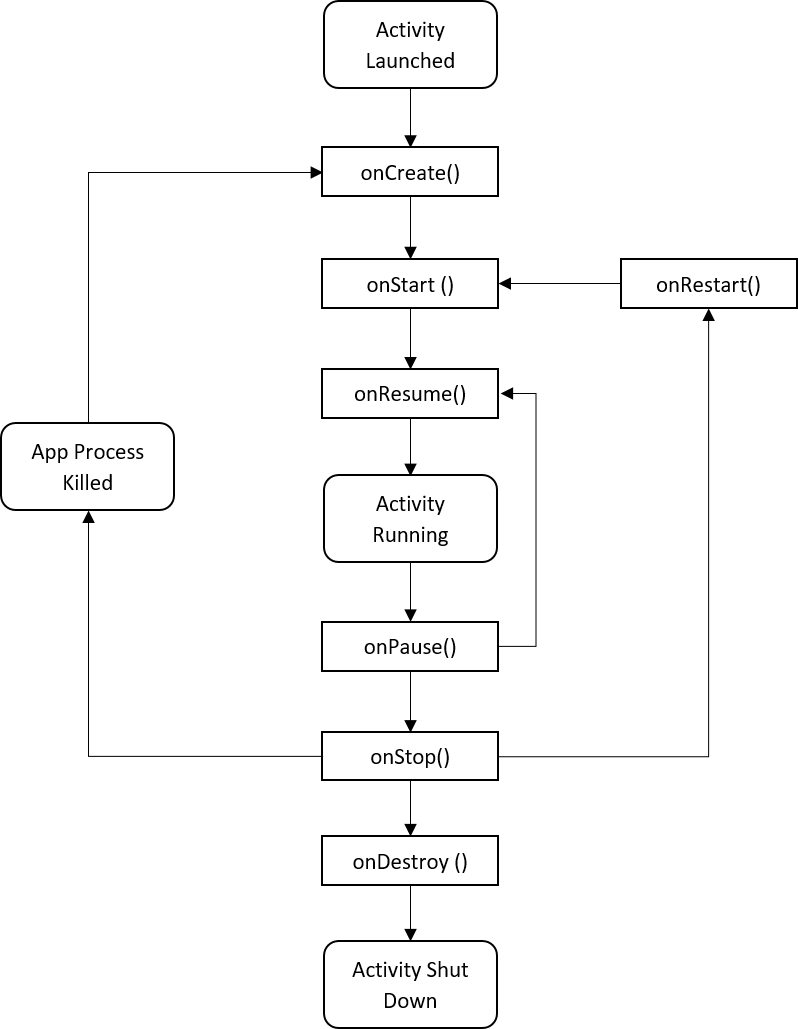}
	\end{center}
	\vspace{-10pt}
	\caption{A simplified illustration of activity lifecycle~\cite{activity}.}
	\label{fig:activity}
\end{figure}
An Android app consists of some activities. An activity manages pages that can be seen on the device screen by users. According to the Android developer documentation, the activity lifecycle is specified by different states that its instances move between them. Each activity class provides a number of callback functions by which the activity instance is informed that its state has changed. For example, the system is creating, stopping, or resuming an activity, or destroying the process in which the activity resides \cite{activity}. In Fig.\ref{fig:activity}, an activity lifecycle is shown. Each activity includes some visual components, which are implemented by the widget package like Button, EditText, and TextView. We call them widgets in this paper. Each widget has a unique ID, which is collected in the R class of an app. With this ID, developers could access supporting widget class methods.

Android offers a mechanism for inter-process communication (IPC) using remote procedure calls, by which a method is called by activity or other application components, but executed remotely (in another process), while its result is returned to the caller \cite{ipc}. With the content provider, it hides the details of how the interprocess communication is managed.

As we mentioned before, Google SDK is an inseparable part of Android apps, which has an event-driven nature. Therefore, in the concolic execution of Android apps, one problem is modeling events. Usually, Android apps are written in Java, but there are fundamental differences between Android and Java programs. Also, we utilize SPF as the concolic execution engine since there is no concolic and even symbolic engine for Android apps. Challenges of testing Android apps with concolic execution are:
\begin{itemize}
	\item Android code is run within DVM\footnote{Dalvik Virtual Machine}. It means that code is compiled to Dalvik bytecode. However, Java code is run within JVM and compiled to Java bytecode. Besides, unlike Java programs, which start from the main function, there is no such function in Android apps. These apps are event-driven, so they can start by tapping an icon or by receiving an SMS and tapping a URL within a text. To use Java engine, we change android code in order to run it within JVM. For this goal, we produce DummyMain classes by static analysis from which the Android programs start. This class simulates the events as the consequence of the user involvement or operating system interaction by calling related functions.
	\item Android apps are too dependent on the SDK. Therefore, it causes path divergence in testing \cite{nariman12}. In symbolic execution, path divergence means the execution of a path leads to call a framework or a library function with symbolic values, making the execution diverge from the developed code. Path divergence causes two main problems:
	\begin{itemize}
		\item It is possible that while testing these libraries or frameworks, functions do not exist. For example, they exist in the device but not in the testing environment.
		\item If we assume that these methods are present, some constraints are added to the path condition due to the conditional statements in their body. Therefore, instead of focusing on developer code, we stuck with testing of the framework or third-party libraries code, which we trust in them.
	\end{itemize}
	
	As a typical example, accessing and modifying data communicated between apps are possible by IPC mechanisms. For example, data stored in SQLite database of an app can be accessed by other apps with this mechanism. To this aim, the first app sends its query to SDK. Then, SDK sends it to the other app. For testing SQL injection vulnerability through IPC, path divergence problem is essential. To overcome this issue, we present the idea of mock classes.
	
\end{itemize}

\subsection{Taint Analysis}	\label{backtaint}
The purpose of taint analysis is to track information flow between sources and sinks \cite{allyouever}. Tainted values are derived from sources, and other values are treated as untainted. For each taint analysis, a policy is defined and enforced. The policy determines which values are tainted, how they propagate through the program, and how they should be analyzed.

In this work, we are going to use dynamic taint analysis in order to detect SQL injection vulnerability in Android apps. In our analysis, sources are the inputs of an app, and sinks are the vulnerable functions. As the source and sink functions belong to SDK classes, so they are used within mock classes. We make the input and output values of these functions symbolic, to denote tainted values, and hence, the resulting classes are called \emph{symbolic mock classes}. Our taint propagation and SQL injection vulnerability detection policy are explained in sections \ref{dynamicTaint} and \ref{sqlidetection}, respectively.

\subsection{SQL Injection in App or through IPC}\label{sqliandroid}
\begin{listing}[b]
	\caption{Secure and insecure way of building query in Android}			\label{sqliCode}					
	\begin{minted}[mathescape,
	linenos,
	fontsize=\scriptsize,
	numbersep=-5pt,
	gobble=2,
	frame=lines,
	tabsize=1,
	breaklines,
	obeytabs=true
	framesep=2mm]{java}
				String st = editText.getText().toString();
				Cursor c = db.rawQuery("SELECT * FROM student WHERE stdno = '"+st+"'", null); //inecure way
				Cursor c = db.rawQuery("SELECT * FROM student WHERE stdno=?'", new String[]{st});//secure way 
				textView.setText(buffer);
	\end{minted}
\end{listing}
Android apps use an SQLite database for storing data. SDK provides some libraries to manage using SQLite. There are secure and insecure methods for implementing database queries in this library. For implementing securely, the developer should use parametric functions in the SDK. In parametric functions, a developer uses ``?" character instead of user inputs. In the database, when a parametric function is used, first the parse tree of the query is built and then the user data substitute for ``?" in the tree. This method prevents injecting commands in queries. In Listing.~\ref{sqliCode}, lines 2 and 3 are an example of the insecure and secure method of building queries, respectively.

Android provides a mechanism for accessing and modifying other app data like SQLite databases if the second app permits. This feature is possible through IPC. If the developer of the second app does not use parametric functions, then SQL injection is possible between the two apps.

List of all functions in SDK through which SQL injection is possible is \{\textit{\textbf{query, queryWithFactory, rawQuery, rawQueryWithFactory, update, updateWithOnConfilict, delete, execSQL}}\}.

For detecting an SQL injection vulnerability, the following conditions should be present in apps: 1) existence of a path from an app input to a vulnerable function, 2) don't use of parametric functions, and finally 3) the result of query propagates to leakage functions by which an attacker can access the result of a database query. For example, printing the result on screen by TextView is considered as a leakage function. In this paper, we focus on the classic form of SQL injection due to our goal in this work is not supporting complicated kind of injections e.g., blind, time-based. 


\begin{figure*}
	\centering
	\includegraphics[scale=0.85]{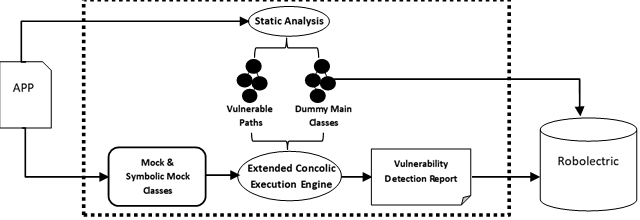}
	\caption{Overview of ConsiDroid.}
	\label{fig:over}       
\end{figure*}
\section{Overview of ConsiDroid}\label{Overview}
An overview of ConsiDroid is given in Fig. \ref{fig:over}. Also, we describe each part with the help of a simple android app (Listing \ref{simpleAppCode}). ConsiDroid has five main parts:
\begin{listing}[b]
	\caption{Simple Vulnerable Android App.}			\label{simpleAppCode}					
	\begin{minted}[mathescape,
	linenos,
	fontsize=\scriptsize,
	numbersep=-5pt,
	gobble=2,
	frame=lines,
	tabsize=2,
	breaklines,
	obeytabs=true
	framesep=2mm]{java}
			public class MainActivity extends Activity{
				protected void onCreate(Bundle savedInstanceState){
					super.onCreate(savedInstanceState);
					EditText et=(EditText)findViewById(R.id.editText);
					final String st=et.getText().toString();
					Button b=(Button)findViewById(R.id.button);
					//some code for inserting data in db.
					b.setOnClickListener(new View.OnClickListener(){
						public void onClick(View v){
							db.rawQuery("SELECT * FROM student where stdno='"+st+"'");
						}
					});
				}
			}
	\end{minted}
\end{listing}
\begin{enumerate}
	\item \textbf{Static Analysis}: By static analysis, we have two primary goals. First is producing the main function in order to compile and run the app in JVM. We produce the main function in a class, which we call it \emph{DummyMain}. We extend each app's code with a set of DummyMain classes, which each of them is a possible execution path in the original app. Second is optimizing our dynamic analysis. To overcome the path explosion problem of concolic execution, we make our analysis hybrid and targeted. In other words, through static analysis, we limit the execution of an app to the desired paths, called \emph{vulnerable paths}, during our concolic execution.
	\item \textbf{Mock and Symbolic Mock Classes}:As we mentioned before, we use SPF. Therefore, for running Android apps on JVM, SDK libraries and its functions should be modeled. We use mock classes for modeling the SDK. Besides, for our taint analysis, we need to track the propagation of tainted variables from source to sink functions. We present symbolic mock classes idea for some specific SDK libraries to complete our taint analysis.
	\item \textbf{Extended Concolic Execution Engine}: Our concolic execution engine is SPF, which is a Java testing tool. With extending Android app code by DummyMain, mock, and symbolic mock classes without changing the original app's code, we can test them on SPF. We extend SPF in two aspects. First, we develop a component for SQL injection vulnerability detection. Second, we manage concolic execution to examine the vulnerable paths extracted by our static analysis at first.
	\begin{listing}[b]
		\caption{Sample Vulnerability Detection Report.}			\label{resCode}					
		\begin{minted}[mathescape,
		linenos,
		fontsize=\scriptsize,
		numbersep=-5pt,
		gobble=2,
		frame=lines,
		tabsize=2,
		breaklines,
		obeytabs=true
		framesep=2mm]{java}
			//STACK TRACE:
			1)android.database.sqlite.SQLiteDatabase. rawQueryWithFactory(SQLiteDatabase$CursorFactory, String,String[],String,CancellationSignal)
			2)android.database.sqlite.SQLiteDatabase. rawQuery(String,String[])
			3)com.example.lab.testak_textinput.MainActivity$. onClick(View)
			4)com.example.lab.testak_textinput.dummyMain. main(String[])
			//APP'S INPUTS THAT CAUSE INJECTION VULNERABILIY:
			1)R.id.editText//developer sanitizer for this input is OFF
			//OBJECT THAT CAUSE LEAKAGE:
			1)android.widget.TextView.setText()	
			//INPUTS OF VULNERABLE FUNCTION
			1)"SELECT * FROM student where stdno='"+st+"'"
		\end{minted}
	\end{listing}
	\item \textbf{Vulnerability Detection Report}:Our analysis results help developers to patch their apps and fix SQL injection vulnerabilities. In the report, we present the ID and the name of the source and sink functions. Also, we expose the stack trace of the program from the source to each vulnerable function. Besides, if parametric functions were not utilized to secure the code, we highlight them in our report. Also, we present the input of the vulnerable function, which helps the analyst and developer to detect and fix the vulnerability. In Listing~\ref{resCode}, the vulnerability report for Listing~\ref{simpleAppCode} is shown with all the above details.
	\item \textbf{Robolectric}: For validating our result, we use Robolectric~\cite{robolectric}, which is a unit testing tool for Android apps. For testing an app with Robolectric, we should specify the testing path. We use DummyMain class and vulnerability detection report to build Robolectric test inputs.
\end{enumerate}

As a summary, before analyzing, we need to change the android app to a java program. To this aim, we extend the android app with DummyMain class and necessary mock and symbolic mock classes. Then our extended concolic execution engine can analyze the program. For example, extending the Android app that is shown in Listing~\ref{simpleAppCode} is done by coupling DummyMain class in Listing ~\ref{dummyCode}. Also, we embed mock classes such as Activity, Bundle, and Button. Besides, we join symbolic mock classes such as EditText (Listing~\ref{mockCode}~(Right)), TextView, and SQLiteDatabase. ConsiDroid analyzes the extended app and produces the vulnerability detection report (Listing~\ref{resCode}). Each item in this section is explained next.

\section{Static Analysis}\label{staticAnalysis}
In our approach, we take advantage of static analysis to produce DummyMain classes. Our smart generation method causes the execution tree of the app to be pruned. Furthermore, we prioritize the paths of the resulting execution tree by static analysis to make dynamic analysis to explore its vulnerable paths at first. We use the Soot\cite{soot} framework to analyze apps statically. Soot extracts the essential graphs that our static analysis relies on. In this part, we discuss our algorithms.

\subsection{Generation of DummyMain classes}\label{dummyMain}
As we mentioned before, unlike Java programs, there is no ``main" function in Android apps. Therefore, for testing an app on SPF and running them on JVM, we need to extend its code. Our extending class is called \emph{DummyMain}, which is produced by using static analysis without any changes in the original app's code. For static analysis, we take advantage of the call graph (CG) of the program, which is built by the Soot framework. This graph is based on Android features and produced by connecting all possible sequences of calling functions to its root node. If we traverse the CG by a DFS algorithm, we can generate all possible DummyMain classes. For optimizing our analysis, we focus on testing extended apps with DummyMain classes, which lead to call one of the vulnerable functions during their execution. We traverse the CG backward from each vulnerable function to the root node, and a DummyManin class is generated from each possible backward path. It worth mentioning that if there is not any vulnerable function in an app, we do not continue its analysis. During the backward traversal of CGs, we collect information about called functions. There are three different types of functions. The first type is Normal, which is built by a developer in an app. Second is Listener that the name of its class contains the special character ``\$". This type of functions is called when an event is produced. Besides, for calling these functions, first, we should call their corresponding parent class that its name is obtained by omitting \$ suffix. For distigushing android event functions from inner classes may define in an app, we take advantage of a list. We produce the list with the name of all possible event functions. The third is Android framework functions, for example, functions which are called in the life cycle of activity. These kinds of functions are called in DummyMain class when we want to call a class extending the Activity class.

\begin{figure*}
	\begin{tabular}{p{0.5\textwidth}p{0.5\textwidth}}
		\begin{minipage}{.50\textwidth}
			\centering
			\includegraphics[scale=0.9, left]{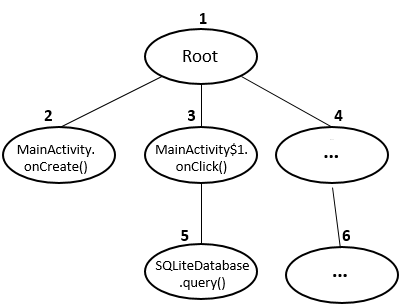}
			\caption{The call graph of a simple app.}
			\label{fig:cg}
		\end{minipage}
		&
		\noindent\begin{minipage}{.44\textwidth}
			\begin{listing}[H]
				\caption{A sample of DummyMain class.}
				\label{dummyCode}
				\begin{minted}[mathescape,
				linenos,
				fontsize=\scriptsize,
				numbersep=-5pt,
				gobble=2,
				frame=lines,
				tabsize=2,
				breaklines,
				obeytabs=true
				framesep=2mm]{java}
			public class DummyMain {
				public static void main(String[] args) {
					MainActivity ma=new MainActivity();
					ma.onCreate(null);
					Button b = (Button)ma.findViewById( R.id.button);
					b.performClick();
				}
			}
				\end{minted}
			\end{listing}
		\end{minipage}
	\end{tabular}
\end{figure*}
As an example, in Fig.~\ref{fig:cg}, the CG of the simple app in Listing.~\ref{simpleAppCode} is shown. This app consists of one activity. Also, there are Button, EditText, and TextView in MainActivity. When Button is clicked, the string fed in EditText is used to generate an SQLite query, and the result of the query is passed to TextView. Only one DummyMain class is generated for analyzing this app, as shown in Listing \ref{dummyCode}. For producing this DummyMain class, we start from node 5, which contains a vulnerable function. By backward traversal, we visit the node 3, which is a listener function and then the root node. For building the code, we start from the last node except the root in our backward traversal, which is node 3. The vulnerable function SQLiteDatabase.query() is called by MainActivity\$1.onClick(). Therefore, we should produce code that causes executing the onClick function. The parent class of the listener function onClick is MainActivity, which extends Activity. As we mentioned before, for listener functions, their parent classes should be called first, resulting in the code at line 3 in Listing~\ref{dummyCode}. Besides, line 4 is added because MainActivity extends Activity. Therefore, functions which are related to the Android framework should be called in order to follow up on the Activity life cycle, as illustrated in Fig.~\ref{fig:activity}. For brevity, the other functions of the life cycle have been omitted here. The listener onClick is performed when an event is produced by the interaction of users with the app. In DummyMain, we simulate this event by the code at lines 5 and 6. By static analysis of the MainActivity code, we find ID (R.id.button) and type (Button) of the relating component with the onClick listener function. Line 6  is the code which simulates the tap event on the Button.

\subsection{Prioritize execution paths}\label{vulPath}

We optimize our dynamic analysis by limiting it to test the execution paths of an extended app with DummyMain classes leading to call vulnerable functions. We improve these tests by forcing SPF to execute our desired paths at first. By this idea, first, we visit nodes in the execution tree of the extended app that are on vulnerable paths. By static analysis, Soot extracts inter-control flow graph (ICFG) of the app, which contains function calls in addition to the control flow graph of each function. We find vulnerable paths from ICFG by backward traversing from a vulnerable function call to the root node. During the traversal, we collect information about each conditional statements. For each conditional branch, we also push the precedence of \emph{then} branch over \emph{else} or vice versa in a stack. We use stack because in concolic execution, unlike our static analysis, we traverse the execution tree in a forward fashion. Therefore, the top entry in the stack is referred to the first conditional statement in the concolic execution.

\begin{figure}[t]
	\begin{center}
		\includegraphics[scale=0.7]{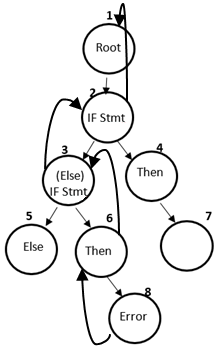}
	\end{center}
	\caption{A sample of an ICFG: the node labeled by ``Error'' denotes a runtime exception statement while an empty node denotes a regular statement.}
	\label{fig:icfg}
\end{figure}

In Fig.~\ref{fig:icfg}, a simple ICFG is shown. Without our static analysis, SPF executes it by a DFS algorithm in which \emph{then} branches have precedence over \emph{else} in conditional statements. Our static analysis extracts the vulnerable path passing through 1,2,3,6 and 8 nodes. In the stack, we push precedence of \emph{then} for the node 3 and after that \emph{else} branch for the node 2. Therefore, after visiting the nodes 1 and 2, we force SPF to execute node 3 and then force it to run the node 6. By this idea, concolic execution analyzes the vulnerable path at first.
\section{Mock and Symbolic Mock Classes}\label{mock}
Mock classes are the same as their corresponding real ones except that the bodies of their functions have been removed. Also,  their return values have been changed to default ones. For example, in Listing \ref{mockCode} (left), a part of EditText's mock class is shown. In this class, the return value of \emph{getText} function is set to \emph{null} at line 7. Symbolic mock classes are the same as mock ones except that symbolic values are returned instead of the default value. In Listing \ref{mockCode} (right) line 7, the return value of \emph{getText} has been made symbolic by \emph{makeSymbolicString}, which is an SPF engine function.
\begin{figure*}[b]
	\noindent\begin{minipage}{.48\textwidth}
		\begin{minted}[mathescape,
		linenos,
		fontsize=\scriptsize,
		numbersep=-5pt,
		gobble=2,
		frame=lines,
		tabsize=1,
		breaklines,
		obeytabs=true
		framesep=2mm]{java}
				public class EditText extends View {
					private String content;
					public EditText(String text) {
						this.content = text;
					}
				public String getText() {	
					return null;
				}
		\end{minted}
	\end{minipage}\hfill
	\begin{minipage}{.48\textwidth}
		\begin{minted}[mathescape,
		linenos,
		fontsize=\scriptsize,
		numbersep=-5pt,
		gobble=2,
		frame=lines,
		tabsize=1,
		breaklines,
		obeytabs=true
		framesep=2mm]{java}
				public class EditText extends View {
					private String content;
					public EditText(String text) {
						this.content = text;
					}
					public String getText() {	
						return makeSymbolicString();
					}
		\end{minted}
	\end{minipage}
\caption{The mock (left) and symbolic mock (right) classes generated for EditText}
\label{mockCode}
\end{figure*}
For running extended Android apps on JVM and preventing path divergence problem, we need to mock SDK classes. Mocking SDK classes need hard effort in order to simulate the Android environment. To this aim, we produce them manually. We produce them concerning the logic of the target app and Android apps life cycle. The idea of mocked classes come from the function summary method \cite{compositional}. The function summary method was introduced to prevent multiple executions of specific functions over different runs. By this method, the body of each function is tested at most once. During the execution of a body, constraints from conditional statements are collected.  In the next function calls, instead of executing the function again, these constraints are added to the path condition. We take advantage of this idea within the mock class idea. In mocking, the default constraints of functions are always ``true". Therefore, they have not any effect on the path conditions of the app.

In addition to mock classes, we also produce symbolic mock for some specific SDK classes, which contain input or vulnerable functions. As we mentioned in Section \ref{backtaint}, we perform our taint analysis in parallel with concolic execution by making tainted values symbolic. Symbolic mock classes help us in this context. It means that we make the input values of an app and also the result of vulnerable functions tainted.

\section{Extended Concolic Execution Engine}\label{extendedConcolic}
For detecting SQL injection vulnerability, we use dynamic taint analysis. For detecting the vulnerability, we should define our security policy of the detection \cite{allyouever}. We define our security policy 
in Sections \ref{dynamicTaint} and \ref{sqlidetection} after discussing our optimization approach in Section \ref{targetedConcolic}. It worth mentioning that for concolic execution, we extend SPF to support targeted concolic execution. Besides, we develop new a component for SPF in order to support SQL injection vulnerability detection.

\subsection{Targeted Concolic Execution}\label{targetedConcolic}
We limit our analysis to extended apps with DummyMain classes, which lead to calling vulnerable functions. Besides, for optimizing concolic execution, we use the vulnerable paths, which are found by our static analysis (see Section~\ref{staticAnalysis}). SPF analyzes the program and traverses the execution tree forward in concolic execution. For each conditional statement in the code, SPF has two choices, e.g., \emph{then} and \emph{else} branches. By default, SPF chooses the \emph{then} branch. Therefore, SPF traverses the tree with a DFS algorithm. We prioritize branches, collected in the stack through the static analysis, which is explained in Section~\ref{vulPath}. By the help of our extension, SPF first analyzes the vulnerable paths. By this idea, we improve the time and memory of analysis.

\subsection{Dynamic Taint Analysis}\label{dynamicTaint}
Injection attacks occur by manipulating the input data of a program and make them malicious. The input of an Android app could be at various points, such as user interface, network, file, system notifications, and IPC. From these points, data enter the app and propagate. If there is a path from input channels to vulnerable functions, there could be a chance for injection vulnerability. Also, a successful injection attack happens when the result of vulnerable functions propagates to leakage functions to be observed by an attacker. Leakage functions are widgets in various points such as user interface, network, file, and IPC.

To reduce the number of false alarms, we use dynamic taint analysis for detecting injection vulnerability. For this goal, we use concolic execution in combination with taint analysis by making tainted values symbolic. SPF can make specific variables symbolic. With SPF and symbolic mocked classes, we perform dynamic taint analysis for Android apps.

\subsection{SQL Injection Detection}\label{sqlidetection}
Apps connect to their SQLite database or may connect to other apps’ database with IPC mechanisms. In both scenarios, there is a possibility of SQL injection vulnerability. We design an algorithm and develop it as an SPF component for detecting SQL injection based on dynamic taint analysis and concolic execution.

In our analysis, we should produce symbolic mock of some classes for tracking the propagation of symbolic values in the program as tainted values. These symbolic mock classes are inputs of the app like EditText or classes containing vulnerable functions like SQLiteDatabase. For supporting SQL injection detection through IPC (see Sections \ref{android} and \ref{sqliandroid}), it is enough to produce a mock class for ContentProvider of SDK by removing all statements except SQLiteDatabase function call statements in the body of its functions. \textit{ContentProvider} is the name of implemented SDK class for content provider concept.

We run the concolic execution until a vulnerable function is called. Next, we check its input argument. If the input contains a symbolic variable (e.g., it is tainted value), then we check if its function development is parametric or not (see section \ref{sqliandroid}). If it is not parametric, there could be a chance of injection vulnerability. Otherwise, it is secure. For completing the chain of a vulnerability occurrence, we continue concolic execution until a leakage function's call. If the input argument of the leakage function is symbolic, which is coming from the result of the vulnerable function, we found a path from an input to a vulnerable function proceeded by a leakage function.

\section{Exploitability Testing by Robolectric}\label{robo}
By dynamic taint analysis, we find all the paths in a program, which conforms with our detection security policy. For ensuring the result of ConsiDroid, we use Robolectric, which is a testing tool for Android apps and independent of the Android environment for its tests. Robolectric needs a target path for testing. In our work, we use DummyMain class (see Listing \ref{dummyCode}) and vulnerability detection report (see Listing \ref{resCode}) as its inputs to generate the Robolectric test input.

\begin{listing}[b]
	\caption{A sample input code of Robolectric for exploiting the app.}
	\label{robocode}
	\begin{minted}[mathescape,
	linenos,
	fontsize=\scriptsize,
	numbersep=-5pt,
	gobble=2,
	frame=lines,
	tabsize=2,
	breaklines,
	obeytabs=true
	framesep=2mm]{java}
			public void SqlInjection_Exploitability() throws Exception {
				Activity ma = Robolectric.setupActivity(MainActivity.class);
				Button b= (Button) ma.findViewById(R.id.button);
				EditText et = (EditText) ma.findViewById(R.id.editText);
				TextView tv = (TextView) ma.findViewById(R.id.textview);
				et.setText("a' or '1'='1");
				b.performClick();
				Logger.error((String) tv.getText(),null);
			}
	\end{minted}
\end{listing}

In Listing~\ref{robocode}, an input of Robolectric is shown. There are many similarities between this code and the code of Listing~\ref{dummyCode}. Lines 4, 5, 6 and 8 are new in this code. The new lines contain IDs of the input (line 4) and leakage (line 5) widgets, which have been collected in the report. For testing SQL injection, we use malicious inputs like “a' or '1'='1” (line 6). For testing other types of SQL injection vulnerability, we should use other input strings, which we can guess them by the query input in the vulnerability detection report. The output of Robolectric execution, which is the result of a malicious query leaked by “tv” object, can prove the existence of vulnerability (line 8).
\section{Evaluation}	\label{evaluation}

To evaluate ConsiDroid, we formulate three research questions:
\begin{enumerate}
	\item Is ConsiDroid capable of generating test cases for real-world Android apps?
	\item How scalable is the approach in detecting SQL injection vulnerabilities for real-world apps?
	\item How well does ConsiDroid perform? Can ConsiDroid detect SQL injection vulnerabilities in a reasonable time? How much code coverage is needed to detect a vulnerability?
\end{enumerate}

In our experiments, we use Ubuntu Linux 16.04 installed on a virtual machine with 12 gigabytes RAM, configured with one processor. This VM is running on a machine with Intel(R) Core(TM) i7-6700 3.4GHz processor. 

\begin{table}[!htbp]
	\footnotesize
	\centering
	\caption{Open-source apps, selected randomly from F-Droid.}
	\label{tab:f-droid}
	\begin{tabular}{*5c}
		\toprule
		{} &\multicolumn{1}{p{1cm}}{\centering Apps}& \multicolumn{1}{p{1.5cm}}{\centering Total Methods} &\multicolumn{1}{p{1.8cm}}{\centering Max Number of Activities}&\multicolumn{1}{p{1.5cm}}{\centering Suspected Apps} \\
		\midrule
		Type-0	&22& $<$ 5000      &9 & 1\\
		Type-1	&10&5000 - 10000   &18& 3\\
		Type-2	&28&10000 - 15000  &9 & 1\\
		Type-3	&31&15000 - 20000  &20& 6\\
		Type-4	&29&20000 - 25000  &21& 1\\
		Type-5	&16&25000 - 30000  &16& 1\\
		Type-6	&4 & $>$ 30000     &7 & 1\\
		\bottomrule
	\end{tabular}
\end{table} 

We choose the apps randomly without any limitation. 

According to apps total number of methods, we have categorized them into seven types (Table \ref{tab:f-droid}). The total number of methods has a direct effect on finding vulnerable methods in each app. To characterize each type of category, we have also measured the maximum number of activities of its apps\footnote{We extracted the total number of methods, SDK methods, and activities for each app with the help of apkanalyzer\cite{apkanalyzer} tool. This tool works on the dexcode of a class or method in smali format.}.
As a result, ConsiDroid generated  DummyMain classes only for $14$ apps (due to the existence of at least one path from a vulnerable function to the root of their CGs) by its static analysis. It means that ConsiDroid can not find any vulnerable functions or suspected paths in other apps. We checked the correctness of our generation by reviewing the apps manually before applying our dynamic analysis. From these apps, five cases can be analyzed by ConsiDroid, and nine of them can not be investigated due to their code obfuscation. As we know, static analysis approaches cannot handle obfuscated code.  
All the five apps were reported by our dynamic analyzer vulnerable to SQL injection except two that were protected by parametric functions. We evaluated the reports by using Robolectric, as explained in Section \ref{robo}, and they were genuinely vulnerable to SQL injection. From these five vulnerable apps, one of them was vulnerable through the IPC mechanism.


For addressing question 2, ConsiDroid can find SQL injection vulnerability in apps with different types (Table~\ref{tab:f-droid}). To show the complexity of these apps, we characterize the $14$ suspected apps in terms of two additional features (Table~\ref{tab:vul-apps}). These features are the number of exploited SDK classes and the maximum method call sequence depth\footnote{We extracted the maximum method call sequence depth with the help of Soot. It worth mentioning that Soot analyzes Android apps in Jimple format, which is a simplified version of Java source code that has a maximum of three components per statement.}. The number of SDK classes specifies the hardness of producing mock classes while the maximum method call sequence depth determines the difficulty of producing DummyMain classes. We also added three vulnerable apps at the end of the table, one known and two manually generated. 
The apps that dynamic analysis was not applied to them (due to the false-positive result by the static analysis) have been specified by $N/A$ (Not Applied) flag in the ``Result'' column. The apps that were reported vulnerable to SQL injection but protected by parametric have been identified by $N$ flag. The apps with $Y$ flag were reported as vulnerable. 

\begin{table*}[t]
	\begin{threeparttable}
		\footnotesize
		\centering
		\caption{Specification of 17 apps, which is analyzed as suspected in static phase.}
		\label{tab:vul-apps}
		\begin{tabular*}{\textwidth}{@{\extracolsep{\stretch{1}}}*{7}{c}@{}}
			\toprule
			Name &Type &Total Methods&SDK Methods& Activity& Max Method Call Sequence& Result
			\\
			\midrule
			clock				&Type-0&1306 &532 & 3&17&N/A\\
			mandarin			&Type-1&6930 &2240&18&21&N/A\\
			ktodo				&Type-1&8716 &2725& 4&13&N/A\\
			silectric			&Type-1&8984 &2488& 5& 6&N/A\\
			musicplayer			&Type-2&13112&3564& 4&10&Y  \\
			simpleaccounting	&Type-3&15086&4290& 6&13&N/A\\
			freetrackgps		&Type-3&16385&3677& 9&16&Y	\\
			iseeu				&Type-3&16670&4052& 2&15&Y	\\
			tinykeepass			&Type-3&16975&4393& 4&18&N/A\\
			Tweetin				&Type-3&18962&3768& 7&12&N/A\\
			smsdroid			&Type-3&19067&4667&12&15& N*\\
			reminder			&Type-4&20287&4626& 3& 7&N	\\
			blackberrymanager	&Type-5&28635&5759& 8&33&N/A\\
			WifiLocationLogger	&Type-6&37017&6375& 1&15&N/A\\
			sieve**				&Type-0&4006 &1210& 8&10&Y	\\
			testak-1***			&Type-4&21533&4991& 1& 5&Y	\\
			testak-2***			&Type-4&21537&4996& 2& 5&Y	\\
			\bottomrule
		\end{tabular*}
		\begin{tablenotes}
			\begin{scriptsize}
				\item * This app used ContentProvider.
				\item ** This app is used by OWASP.
				\item *** These apps are developed by author.
			\end{scriptsize}
		\end{tablenotes}
	\end{threeparttable}
\end{table*}
\begin{table*}[t]
	\footnotesize
	\centering
	\caption{Characterization of apps in terms of the percentile rank of their complexity metrics. The number of produced DummyMain classes, the amount of time and code coverage needed to detect SQL injection vulnerability are present.}
	\label{tab:ConsiDroid-time-cov}
	\begin{tabular*}{\textwidth}{@{\extracolsep{\stretch{1}}}*{8}{c}@{}}
		\toprule
		Name &  No. M &  No. SDK &  No. Act & Max MCS &Time(ms) & Code Coverage & Produced DummyMains \\
		\midrule
		musicplayer & 34\% & 34\% & 58\% & 34\%  & 49      & 52\%  &  3  \\
		freetrackgps& 45\% & 37\% & 90\% & 66\%  & 22      & 33\%  &  8  \\
		iseeu      	& 49\% & 45\% & 21\% & 60\%  & 13      & 24\%  &  2  \\
		smsdroid 	& 61\% & 64\% & 95\% & 59\%  & 23      & 36\%  &  2  \\
		reminder 	& 66\% & 62\% & 40\% & 19\%  & 17      & 26\%  &  3  \\
		sieve 		& 12\% & 12\% & 86\% & 33\%  & 24      & 31\%  &  4  \\
		testak-1 	& 72\% & 71\% & 2\%  & 4\%   & 41      & 46\%  &  2  \\
		testak-2 	& 73\% & 72\% & 22\% & 6\%   & 37      & 44\%  &  3  \\
		\bottomrule
	\end{tabular*}
\end{table*}


Our dynamic analysis was applied to eight apps. We measure the complexity of them by four metrics, which examine how apps' complexity affects the execution time of the dynamic analysis. Also, these metrics study the scalability of our approach. These metrics are the total number of methods (No. M), the number of exploited SDK classes (No. SDK),  the maximum number of activities (No. Act), and the maximum method call sequence depth (Max MCS). To this aim, we computed the percentile rank of their metrics, as shown in Table \ref{tab:ConsiDroid-time-cov}. Following the approach of \cite{sigdroid}, the complexity class of each app can be computed in terms of the percentile rank of their metrics. An application belongs to the $10$th overall complexity class if it belongs to the $10$th percentile in the four dimensions. In other words, an app belonging to a lower class is less complicated concerning all four dimensions compared to an app from a higher class. As it is illustrated in this table, ``smsdriod'' is the most sophisticated app among them. The increase in the app complexity results in a small increase in our dynamic analysis execution time due to our targeted analysis. We can conclude that ConsiDroid is capable of scaling to even the most complex Android apps.

For answering question 3, we present the time and the amount of the code coverage of analyzing each app with ConsiDroid for detection of SQL injection vulnerability in table \ref{tab:ConsiDroid-time-cov}. Besides, we have shown the number of produced DummyMain classes for each app.  With ConsiDroid, we could find the vulnerability almost with less than $50\%$ of the code coverage as a result of our targeted analysis in a reasonable time. As we mentioned before, we present the first tool in this community for finding a vulnerability in Android apps with concolic execution technique. So, there is not any similar tool for comparison with ConsiDroid.



\section{Related Work}	\label{relatedWork}
DART \cite{dart} is the first work that presented concolic execution method for testing programs. KLEE \cite{klee} is another tool for concolic execution. In this work unlike DART, in conditional statements, both branches are executed in parallel for enhancing the time of testing. DART and KLEE are just for C programs. Also, there are some other tool like SAGE \cite{sage}, AEG \cite{aeg} and Mayhem \cite{mayhem} for detecting software vulnerabilities. Besides, to detecting vulnerabilities, AEG and Mayhem produce exploit code automatically. These works support Windows or UNIX based operating systems.

ACTEVE \cite{acteve} is the first paper on testing Android apps with concolic execution. This tool only supports tap event sequences with a maximum length of four. 
Condroid \cite{condroid}, an extension of ACTEVE, detects logic bomb Android malware. 
AppIntent \cite{appintent} is a tool for detecting privacy violation in Android apps with concolic execution. AppIntent uses static taint analysis in order to enhance concolic execution and make a targeted concolic analysis. Malton \cite{malton} is a tool for detecting Android malware apps by using binary analysis with the assistance of Valgrind \cite{valgrind}. Malton analyzes the app inside the device, so there is no need for producing mock classes. Sig-Droid \cite{sigdroid} is a tool for testing Android apps with symbolic execution. We were inspired by this tool for producing mock classes of SDK and using SPF. In paper \cite{ceture} a tool is introduced for detecting Android framework vulnerabilities with symbolic execution. The target of this tool is the Android framework and not Android apps. To the best of our knowledge, our research is the first one for detecting SQL injection vulnerability in Android apps.

In table \ref{tab:android-testing}, there is a comparison between ConsiDroid and other Android testing tools with different dynamic approaches. As it is shown, this comparison is based on searching methods, supporting event types, if they exploit static analysis to enhance their dynamic approach, and the path explosion problem in the tools, which is not applicable to Monkey.

\begin{table*}[t]
	\footnotesize
	\centering
	\caption{Comparison of ConsiDroid with other dynamic testing Android tools. }
	\label{tab:android-testing}
	\begin{tabular*}{\textwidth}{@{\extracolsep{\stretch{1}}}*{5}{c}@{}}
		\toprule
		Tool &  Search method & Events & Static + Dynamic Analysis & Path explosion problem \\
		\midrule
		Monkey      &Random     & Text, System, GUI& No   & -    \\
		ACTEVE      &Concolic	& GUI(Tap Event)   & No   & Yes	 \\
		Sig-Droid   &Symbolic   & Text, GUI		   & No   & Yes	 \\
		ConsiDroid  &Concolic   & Text, System, GUI& Yes  & No	 \\
		\bottomrule
	\end{tabular*}
\end{table*}

Furthermore, we have compared ConsiDroid with concolic or symbolic execution-based tools in table \ref{tab:sec-tool}.Our comparison is based on features which are supporting event-driven nature of Android apps, the path explosion problem in the tools, utilizing static analysis in concolic or symbolic execution, and security related issues such as logic bomb, SQL injection vulnerability, and privacy violation detection.
\begin{table*}[t]
	\footnotesize
	\centering
	\caption{Comparison of ConsiDroid with similar security concolic- or symbolic-based tools.}
	\label{tab:sec-tool}
	\begin{tabular*}{\textwidth}{@{\extracolsep{\stretch{1}}}*{7}{c}@{}}
		\toprule
		Tool &  Event-Driven & Path explosion & Static + Dynamic Analysis & Logic Bomb Detection&\multicolumn{1}{p{3cm}}{\centering SQL Injection Vulnerability Detection}&\multicolumn{1}{p{3cm}}{\centering Privacy Violation Detection} \\
		\midrule
		AppIntent   &Yes&No &Yes&No &No &Yes     \\
		Condroid    &Yes&Yes&No &Yes&No &No 	 \\
		Sig-Droid   &Yes&Yes&No &No &No &No 	 \\
		ConsiDroid  &Yes&No	&Yes&No &Yes&No 	 \\
		\bottomrule
	\end{tabular*}
\end{table*}

Furthermore, there exist tools for identifying a different kind of information leakage in android apps. These tools are based on static and dynamic taint analysis. Flowdroid\cite{soot} is a static taint analysis tool, which we use it to extract CG and CFG of the app in order to produce DummyMain class. TaintDroid\cite{taintdroid} is a dynamic taint analysis tool for capturing information leakages in android apps, which its dynamic approach is not concolic or symbolic. Shahriar et al.\cite{kld-based} presenting KLD-based detection of content leakage vulnerabilities based on three secure programming principles. Their approach focused on ContentProvider leakage, which ConsiDroid can detect them with the symbolic mock class idea. Demissie et al.\cite{AWiDe} present a dynamic and a static taint analysis approach for detecting Android Wicked Delegation, which is based on IPC mechanism.

\section{Discussion}\label{discussion}
Although ConsiDroid is the first Android apps vulnerability detection tool, there are several open problems which need to be dealt with in future work and improvements. ConsiDroid only can detect classic SQL injection vulnerability. There are other complex types of SQL injection like blind SQL injection, which we do not support. For other types of SQL injection, we need to study a different type of leakage functions. Also, there are other kinds of vulnerabilities, which ConsiDroid cannot analyze them. For detecting other kinds of injections such as OS Shell injection, it is required to study vulnerable functions and produce appropriate symbolic mock classes. Besides, the vulnerability detection policy should be modified, and SPF component adjusted accordingly. However, for another kind of vulnerabilities, it needs more research and study. Besides, ConsiDroid is a vulnerability detection tool and not an automatic exploit generation tool. It needs hard effort to enhance ConsiDroid for automatically generate exploits.    

ConsiDroid could only analyze Java coded Android apps. There are apps, which are developed with native or combination of Java and native code. Analyzing these kinds of apps is another direction in future works. ConsiDroid takes advantages of SPF and mocking technique for concolic execution. We generate mock classes manually, which is a time-consuming procedure. As future work, we suggest automating it by extending the code of Robolectric.

Furthermore, ConsiDroid could be enhanced by modeling SDK and Android environment following the same approach as \cite{heila15envmodel} and \cite{heila15libmodel}. Another idea could be presenting an Android concolic engine, which works on Dalvik bytecode or ARM binaries on emulators. With this idea, we do not need mock classes anymore.  With this idea, we can analyze obfuscated apps, which ConsiDroid can not analyze them. It worth mentioning that our static analysis for producing DummyMain classes are not accurate. ConsiDroid could be improved in the future by mapping the nodes of CG to ICFG to compute vulnerable paths more precisely.

For evaluating ConsiDroid, we take advantage of F-Droid\cite{fdroid} repository apps, which are real-world apps without any tags. There exist a dataset called DroidBench\cite{droidbench}, which consist of a bunch of apps for testing taint analysis. Problem with DroidBench is that it does not support a different kind of vulnerabilities like SQL injection. In future, we can extend this repository for supporting vulnerabilities. 
\section{Conclusion}\label{conclusion}
In this paper, we presented a new technique for detecting SQL injection vulnerability in Android apps and a new tool which we called ConsiDroid. In our technique, we managed the lack of Android concolic execution engine by taking advantage of SPF, which is a Java concolic execution engine. For concolic execution of suspected paths in Android apps with SPF, we produced DummyMain classes by backward traversal of app's call graph. Also, we optimized the analysis by giving precedence to the vulnerable paths in concolic execution. For detecting SQL injection vulnerability in Android apps, we combined concolic execution with dynamic taint analysis by symbolic mock classes idea. To achieve this aim, we extended SPF accordingly. As a result, we provided useful information for patching the vulnerabilities in the report. We evaluate ConsiDroid with real-world and open source apps of F-Droid repository. We select 140 apps randomly. From selected apps, ConsiDroid could detect three vulnerable apps accurately. Also, we examine the performance of ConsiDroid based on maximum method call sequence, the number of methods, SDK methods, and activities. Our experiment shows that increasing complexity of apps has a small effect on time and code coverage of test apps.

\section*{Acknowledgments}
We would like to thank Hamid Bagheri, co-author of Sig-Droid\cite{sigdroid}, for his valuable  comments on the paper.
\bibliography{myRef}

\end{document}